\documentclass{aastex}
\usepackage{amsmath,amssymb}  
\begin{document}
\title{Scaling of Black Hole Accretion Discs: From Gamma-Ray Bursts and
Black Hole X-Ray Binaries to Active Galactic Nuclei}
\shorttitle{Scaling of Black Hole Accretion Discs}
\author{J. I. Katz}
\affil{Department of Physics and McDonnell Center for the
Space Sciences \\ Washington University, St. Louis, Mo. 63130}
\email{katz@wuphys.wustl.edu}
\begin{abstract}
I consider how physical processes scale over eight orders of magnitude in
black hole mass, from stellar masses in gamma-ray bursts (GRB) and
black-hole X-ray binaries (BHXRB) to supermassive active galactic nuclei
(AGN).  Accretion rates onto stellar mass black holes range over more than
sixteen orders of magnitude, from the lower luminosity BHXRB to GRB.  These
enormous parameter ranges correspond to qualitative as well as quantitative
differences in behavior.  The fundamental questions involve the balance
between nonequilibrium and thermalized plasmas.  When energy fluxes exceed
a critical value $\sim 10^{29}$ erg/cm$^2$s, as in GRB, a black-body
equilibrium pair plasma forms.  At the lower fluxes found in AGN, BHXRB and
microquasars, accretion power electrodynamically accelerates a small number
of very energetic particles, explaining their non-thermal spectra and the
high energy gamma-ray emission of blazars.  Particle acceleration,
(particularly of leptons because of the energy dependence of the
Klein-Nishina cross-section) is limited by the flux of soft thermal photons,
which is necessarily model-dependent.  This prohibits the formulation of
general scaling laws.  Ultra-high energy cosmic rays
may be accelerated by massive black holes, otherwise undetectable, with very
low thermal luminosities.  New-born fast high-field pulsars may be in the
black-body equilibrium regime, resembling SGR in permanent outburst.  I also
consider the question, significant for the acceleration of nonthermal
particles in GRB outflows, of whether collisionless plasmas interpenetrate
rather than forming hydrodynamic shocks, and propose this as an alternative
to internal shock models of GRB.  A new appendix attempts to explain why
AGN are, proportionally, more efficient accelerators of energetic particles
than stellar mass black holes.
\end{abstract}
\section{Introduction}
\label{introduction}
Hundreds of models of active galactic nuclei (AGN) and quasars have been
published since the pioneering work of \cite{S64}, yet none of them is
completely satisfactory and generally accepted.  This paper\footnote{The
present paper is the ``forthcoming work'' cited as Katz (1997b) by
\cite{K97}.} investigates the general scaling laws that connect the very
different parameter regimes appropriate to AGN and to BHXRB, including
microquasars.

GRB, AGN and BHXRB have certain qualitative similarities, despite enormous 
quantitative differences in their luminosities, masses, and time scales.  
They all are powered by accretion onto a central black hole.  They all
produce nonthermal radiation.  GRB and AGN and some BHXRB show evidence for
relativistic bulk motion.  They all fluctuate irregularly in intensity.
AGN and some BHXRB (microquasars) are directly observed to produce narrow
jets, while energetic arguments and modeling of their afterglows
by \cite{F01} indicate that GRB are also strongly collimated.

These similarities suggest that it may be useful to investigate how a common
accretion disc model may manifest itself differently in these objects.  The
present discussion builds upon the \cite{B76,L76} model for AGN, although
these authors assumed (probably unnecessarily, except to make the problem
tractable) that the magnetic dipole moment is aligned with a rotational
axis.  These models may be scaled to stellar mass BHXRB such as Cyg X-1 and
the microquasars.

A single parameter divides all accretion discs into two classes.  The
fundamental difference between these classes of models is that in GRB and in
SGR in outburst (and probably also in young fast high-field pulsars, as yet
unobserved) the relativistic wind is thermalized to an equilibrium pair
plasma, while at the lower power densities of AGN, BHXRB and observed radio
pulsars the wind remains transparent and very far from thermodynamic
equilibrium with the radiation field, as a comparatively few particles are
accelerated but to high energies.  The condition for thermalization
\cite{K82,K94a,K96,K97} is that 
\begin{equation}
{\cal I} \equiv I/I_{char} \gtrsim 1,
\label{idef}
\end{equation}
where $I$ is the energy flux (power emitted per unit area, whether carried
by particles, energetic photons, or low frequency or even DC Poynting flux),
so that $I$ exceeds a characteristic value
\begin{equation}
I_{char} \equiv \sigma_{SB} \left[{m_e c^2 \over k_B \ln{(6 (M/M_\odot)}
N_{B\odot} \alpha_G \alpha^2)}\right]^4 \sim {4 \times 10^{29} {\rm erg/cm^2
s} \over \left(1 + 0.05\ln{M/M_\odot}\right)^4},
\label{ich}
\end{equation}
where $\sigma_{SB}$ is the Stefan-Boltzmann constant, $N_{B\odot} \approx
1.2 \times 10^{57}$ is the number of baryons in a Solar mass, $\alpha_G
\equiv G m_e^2/\hbar c \approx 1.76 \times 10^{-45}$ is the ``gravitational
fine structure constant'' defined for the electron mass.   This corresponds
to an equivalent black body temperature exceeding the characteristic value
\begin{equation}
T_{ch} \approx {2.9 \times 10^{8\,\circ}K \over 1 + 0.05 \ln{M/M_\odot}}
\label{tch}
\end{equation}
at which a black body equilibrium pair plasma is opaque to Thomson
scattering over the characteristic length scale (three Schwarzschild radii)
of the source.  Because of the steep Boltzmann factor in the equilibrium
pair density, the characteristic temperature and intensity are almost
universal constants, only weakly dependent on the mass of the source, even
over the range from $1 M_\odot$ to $10^8 M_\odot$.

The thermalization of energetic particles by scattering in regions of high
energy density is auto-catalytic: scattering increases the number density of
potential scatterers (by processes such as double Compton scattering, pair
production, and curvature radiation).  However, these processes only run
away to full thermalization if ${\cal I} \gg 1$.

In these models of AGN and BHXRB the electromagnetic energy is converted to
the energy of accelerated particles close to the black hole.  This is in
contrast to GRB in which electromagnetic energy is converted to a black-body
equilibrium pair plasma \cite{PSN93}.

In another class of models of AGN and BHXRB the disc radiates vacuum
electromagnetic waves instead of energetic particles.  At much greater radii
these waves accelerate particles, just as in GRB the pair plasma accelerates
particles in distant collective interactions.  These two classes of models
can be comparably efficient particle accelerators.  I do not consider the
vacuum wave model further because it is less closely analogous to the GRB
model (which cannot be a vacuum wave model because the energy density leads
to creation of an equilibrium pair plasma), and because external plasma
injection or pair breakdown are likely to fill the wave zone with energetic
particles.

The remainder of this paper is chiefly concerned with scaling of a variety of
physical processes over the very wide range of parameters encountered.  It
would be difficult to say anything qualitatively new about any of the models
of these objects, which have been extensively developed over 45 years, but
it may be useful to re-examine their scaling.  To this end it is necessary
to review the fundamental physical processes involved. 

There are also temporal similarities among the various classes of black hole
accretion discs.  In one BHXRB (Cyg X-1; \cite{WSKC78}) a non-zero time
skewness was measured in an X-ray time series.  Time skewness of the same
sense is found (\cite{N94}) in many GRB.  Searches for time skewness in AGN
time series have so far been unsuccessful (\cite{PR97}), but do not exclude
it.  In fact, it was once suggested (\cite{PS75,PS77}) that Cyg X-1 might be
a source of GRB.  Although such Galactic stellar mass X-ray sources are now
known not to be the origin of GRB, Cyg X-1 does show outbursts and flaring
behavior in comparable energy bands (\cite{Z02}).

\section{Homopolar Generators}

All of the objects we discuss contain rotating, electrically conducting,
magnetized matter, and therefore are homopolar generators with potentials
$V_0 \sim \int (v/c) B\,dr$.  In general, such generators produce a power
$P = V_0^2 / Z$, where the load impedance $Z \equiv E/H = Z_r + i Z_i$ is
written as the sum of reactive ($Z_i$) and dissipative ($Z_r$) parts.  A
fundamental assumption of nearly all astrophysical models of such objects is
that $\vert Z \vert \sim Z_0 \equiv 4 \pi / c = 4.19 \times 10^{-10}$
sec/cm, the impedance of free space in c.g.s.~units ($Z_0 \equiv \sqrt{\mu_0
/ \epsilon_0} = 377\ \Omega$ in m.k.s.~units).

The detailed or microscopic justification of $\vert Z \vert \sim Z_0$ is
model-dependent.  A similar result will always be found if the particle
energies are large enough that particle multiplication (by processes such as
double Compton scattering and collisional and radiative pair production)
turn the insulating vacuum into a conducting relativistic plasma.  $Z_i \sim
Z_0$ is a general result obtainable from Maxwell's equations by estimating
$\vert {\vec \nabla} \vert \sim \partial / \partial (ct)$.  As long as the
system size exceeds $c \Delta t$, where $\Delta t$ is a characteristic
(rotation or fluctuation) time scale, the reactive impedance $Z_i\sim Z_0$
will always be in series with $Z_r$.

The dissipative impedance $Z_r$ may be small in a relativistic plasma, but
we hypothesize that when the fields are large particle acceleration and
multiplication will generally lead to $Z_r \sim Z_0$.  This hypothesis,
essentially dimensional, is central to the model of this paper and is
implicit in most pulsar and AGN models (\cite{GJ69,B76,L76}).  For example,
application to pulsars gives
\begin{equation}
P \sim {V^2 \over Z} \sim {c \over 4 \pi} (Er)^2 \sim {c \over 4 \pi}
\left({v \over c} rB\right)^2 \sim {\omega^2 r^4 B^2 \over 4 \pi c} \sim
{\omega^2 \mu^2 \over 4 \pi c r^2} \sim {\omega^4 \mu^2 \over 4 \pi c^3},
\label{gjpower}
\end{equation}
where $\mu$ is the magnetic moment and $B \sim \mu/r^3$ is evaluated at the
radius of light cylinder $r = c/\omega$.  This is comparable both to the
expression for magnetic dipole radiation of a misaligned rotor in vacuum
and to the result of \cite{GJ69} for an aligned rotor when the vacuum
fills with a relativistic plasma.  It is also unavoidable on dimensional
grounds, provided that the fields are large enough to drive these processes.

In the laboratory $Z_r \gg Z_0$ when the vacuum or materials do not break
down into a conducting plasma, as when fields and potentials are small and
insulators remain insulators, and frequently $Z_r \ll Z_0$ when conductors
are not relativistic.  In relativistic astrophysics $Z_r \sim Z_0$ follows
from assuming that the power carried to $r > c \Delta t$ by the Poynting
vector is dissipated somewhere, rather than remaining forever as
electromagnetic field energy or returning (like an unphysical advanced time
solution) to the source.

The first nontrivial part of this problem is to estimate $B$, which can be
done on the basis of dimensional arguments from the total disc power.  The
second, and most critical (but generally unsolved), part is to determine how
the potential drop is distributed in space and how the dissipated power is
apportioned among the accelerated particles. 

\section{Nonthermal efficiency}
\label{nonthermal}

All three classes (AGN, BHXRB and GRB) of objects show a significant amount
of nonthermal emission.  The nonthermal efficiency, which may be defined as
the fraction of the emitted power appearing as non-Planckian radiation or as
particle distributions that are far from thermal equilibrium (typically
power laws over orders of magnitude in energy) is ${\cal O}(1)$.  In GRB 
the observed emission appears to be nonthermal, although it is not the
primary radiation emitted by the central engine but rather the consequence
of particles accelerated in the relativistic outflow; the high energy
density and intensity at the source thermalizes the relativistic wind.
Still, the wind is generally believed to be produced by a fundamentally
nonthermal process, the coherent radiation of electromagnetic energy by the
fields of the central engine.  In many AGN a substantial fraction of the
power appears as nonthermal visible synchrotron radiation or high energy
gamma-rays \cite{K99}.

The case for nonthermal X-ray emission in BHXRB is plausible but less
compelling.  It is a natural explanation of the complex multi-component
spectra frequently observed, but it may also be possible to explain such
spectra as the sum of thermal spectra emitted by matter distributed over a
range of temperatures.  For example, \cite{LB69} gives the classic example of
an apparently non-thermal power law spectrum that results from thermal
emission from an accretion disc whose temperature is a power law function of
radius.  Only if the emission extends to very high energy, or is associated
with low frequency emission of very high brightness temperature (as in
radio pulsars), can the case for nonthermal processes be considered
compelling.

As examples of such a compelling case, the superluminal radio components and
jets present in some BHXRB (microquasars) certainly require acceleration of
relativistic particles.  Some, such as Cyg X-3, also show strong outbursts
of nonthermal radio emission (there is no direct evidence this particular
object contains a black hole, and qualitatively similar effects may be
produced by accretion onto a low magnetic field neutron star), but this
radio emission is only a very small fraction of their power output.

In GRB the electrodynamic efficiency $\epsilon_e$ (Equation 5 of \cite{K97})
is not directly measured because virtually all the thermal radiation emerges
as neutrinos (the thermal luminosity is subject to an Eddington limit that
is $\sim 10^{38} M/M_\odot$ erg/s for electromagnetic radiation but many
orders of magnitude greater for neutrinos because of the tiny neutrino
opacity of matter), and is essentially undetectable.  For the lower energy
density accretion flows of AGN and BHXRB neutrino emission is negligible, and
the thermal radiation produced by viscous heating is directly observable.

\cite{K97} argued that $\epsilon_e \sim 0.1$--0.5 is likely, independent 
of the magnitude of the magnetic field (but depending on its unknown 
orientation and spatial structure).  This is consistent with observations of
AGN; the likelihood of relativistic beaming precludes quantitative 
comparisons.  This range of $\epsilon_e$ is also consistent with 
observations of Cyg X-1 and other BHXRB if the harder components of their 
spectra are either nonthermal or the thermal emission of optically thin 
matter heated by nonthermal particles.

The measured nonthermal efficiency is also affected by radiation trapping
(\cite{K77}).  As is well known (and the subject of an extensive literature)
if the mass accretion rate exceeds the nominal Eddington rate the excess
mass is readily swallowed by the black hole, but the luminosity $L_{th}$
emergent in thermal radiation that diffuses through the accretion flow is
limited to slightly less than the Eddington limiting luminosity $L_E$ ({\it
cf.\/} \cite{ECK88}).  An analogous limit involving the neutrino Eddington
limit applies to the unobserved neutrino luminosity.  In AGN and BHXRB there
can be an apparent nonthermal efficiency $\epsilon_e \to 1$ and $P_{rw} \gg 
L_{th}$, because the nonthermal wind luminosity $P_{rw}$ is proportional to
the accretion rate and is not subject to the Eddington limit, while the
emergent thermal luminosity $L_{th}$ cannot exceed $L_E$ even if the viscous
power dissipated $P_{visc} \gg L_E$.

\section{Particle acceleration}
\label{particle}

In this section I collect and summarize results (many of which are spread
throughout the very extensive literature) concerning nonthermal acceleration
processes that may occur in black hole accretion discs.  The purpose is to
develop criteria to determine if the condition ${\cal I} \gg 1$ {\it
necessarily\/} leads to black-body equilibration, which is energetically
allowed, but not required by any fundamental physical law.  Unfortunately,
the rates of many processes depend sensitively on the angular, spatial and
(in the Klein-Nishina case) spectral distributions of any soft photon
background flux, such as may be emitted by clouds of comparatively cool
thermal matter.  These fluxes depend on detailed modeling that cannot be
developed from first principles, and for which we do not generally have 
sufficient information to enable even phenomenological models.

Pulsars and GRB (in the present model) have large electric fields which
lead to pair production.  A supermassive black hole in a galactic nucleus, 
or a stellar mass black hole in a mass transfer binary, is surrounded by a 
complex accretional gas flow.  Significant sources of mass include the 
companion star (in the BHXRB), the galactic interstellar medium and
possibly disrupted stars (in the AGN) and the surface of the outer parts of
the accretion disc.  Although the flow is not understood in detail, it is 
plausible that some of this gas has sufficiently little angular momentum (or
loses its angular momentum at large enough radii) to permit accretion on the
axis of rotation, and can fill all directions around the black hole, as was
found in the calculations of \cite{ECK88}.  A pulsar-like vacuum is not
likely.  The space charge density required to neutralize the corotational
electric field (\cite{GJ69}) is small, and may readily be supplied by this
plasma.  Pair production is therefore not required for the extraction of
energy from the rotating disc in AGN and BHXRB, although it may occur.

The homopolar generator of a rotating magnetized fluid (\cite{GJ69,B76,L76})
implies a large electric field with a nonzero curl in an inertial
(observer's) frame.  This cannot be canceled or screened {\it everywhere}
by space charge density.  In some regions the electric field must be nonzero
(and large); these regions are conventionally called ``gaps'' because the
field purges them of plasma.  Charged particles entering or created
(\cite{S71,RS75}) in these gaps are accelerated.  The fundamental unsolved
problems of compact magnetized rotating astronomical objects are the
distribution of the homopolar potential drop and the acceleration of
particles in gaps.

If the entire circuit were composed of dense thermal plasma then $Z_r$
would be the plasma resistivity, which would be $\ll Z_0$.  $Z_i$ would
depend on the circuit dimensions.  If these were much less than the
characteristic light travel dimension for time variations (as often the 
case in the laboratory, where circuits of cm or m dimensions may oscillate
at 60 Hz, or even be nearly DC) then $Z_i$ may be tiny ($\gg 1\,\Omega$).
The power $P = V_0^2 Z_r / \vert Z \vert^2 \approx V_0^2/Z_r$ could be
extremely high.  This may describe some GRB (perhaps short GRB or subpulses
in longer GRB, though it is hard to be sure because their magnetic fields
are unknown), in which vacuum gaps may fill with equilibrium pair plasma, or
electromagnetically driven supernovae.  However, if the circuits close at
the light travel distance or are open beyond it, as in classical pulsar 
models \cite{GJ69}, then $Z_i \sim Z_0$ and $P \lesssim V_0^2/Z_0$ (as in
Eq.~\ref{gjpower}).  This is probably the case for AGN and BHXRS, in which
$V_0$ is largely determined by fields near the light travel distance.

Near a luminous object accelerated electrons and positrons are slowed by
Compton scattering on the thermal radiation field (\cite{J65,KS74}).  As a
result, they are not accelerated to the limiting energy $E_0 \equiv eV_0
\sim e \int (v/c) B\,dr$.  As an extreme bound we assume that this electric
field is parallel to $\vec B$.  In the presence of a significant
(\cite{GJ69}) density of charged particles the parallel component of $\vec
E$ will be much less.  The following results will remain as upper bounds
and may be useful as such, even though they are likely to be overestimates
of the actual energies achieved.

\subsection{Electron Retardation by Compton Scattering; Thomson Limit}
\label{thomson}

If the Thomson cross-section is applicable the energy loss length $\ell_C$
of an electron with Lorentz factor $\gamma$ in an isotropic radiation field
of intensity $L_{th} / (4 \pi r^2)$ near a mass $M$ is
\begin{equation}
\ell_C = {m_e \over m_p} {1 \over \gamma} {L_E \over L_{th}} {r c^2 \over
GM} r.
\label{lc}
\end{equation}
The first two factors are each $\ll 1$, and the next two are not much 
greater than unity in the inner disc of a luminous object, so that $\ell_C 
\ll r$.  This result is also approximately valid for anisotropic radiation 
fields, except in the extreme case of a particle moving accurately in the 
direction of a narrowly collimated beam of radiation.

Given a value for the magnetic field $B$, it is possible to calculate the 
maximum energy an electron achieves, and to estimate its radiation.  To
estimate $B$ equate the magnetic stress to that required to supply a
bolometric luminosity $L_b$ (including relativistic wind, thermal radiation
and radiation advected into the black hole) yields
\begin{equation}
{B^2 \over 8 \pi} \sim {1 \over 2} {L_b \over L_E} \left({GM \over r c^2}
\right)^{3/2} {r \over h} {c^4 \over GM \kappa} \sim 3 \times 10^7\ {L_b 
\over L_E} \left({10 GM \over r c^2}\right)^{3/2} {r \over 10 h} {10^8 
M_\odot \over M}\ {{\rm erg} \over {\rm cm}^3},
\label{b}
\end{equation}
where $\kappa$ is the Thomson scattering opacity and $h$ the disc thickness.
We assume only that the viscosity is magnetic and that $B^2 \sim \langle B_r
B_\phi \rangle$.  The magnetic field does not depend on an assumption of
equipartition or on the value of $\alpha$, and is derived from the accretion
rate implied by $L_b$ alone.  This estimate of $B$ is supported by 
polarimetric observation of the BXHS Cyg X-1 (\cite{G03}).

Although the estimate of $B$ (and especially
of $\langle B_r B_\phi \rangle$) is robust, the process by which accretional
energy is dissipated remains mysterious.  It is plausible, or even likely 
(in analogy with the magnetic heating of the Solar convective zone or the
tidal dissipation in the terrestrial oceans),
that dissipation occurs in a disc corona rather than in its deep interior.
In a low density corona the power dissipated may be coupled to a few
particles accelerated to high energy while power dissipated in the deep
interior heats matter in thermal equilibrium.  Depending on the (unknown)
spatial distribution of impedance and magnetic field, these loads combine in
a complex and unpredictable manner.  For example, dense thermal plasma of
low impedance may be in series with or in parallel to vacuum gaps; in the 
former case most of the dissipation occurs in the gaps, while in the latter
case most of it occurs in the thermal plasma.

The tendency of large electric fields to purge space of plasma, either by
collisional resistive heating and subsequent expansion or by acceleration of
collisionless particles, combined with the pinch effect that concentrates
the current into narrow filaments, suggests that there will be regions of
near-vacuum in which a few particles are accelerated to high energy.  Hence
we must consider the limits on particle acceleration in those regions.

Equating the energy $e E \ell_C$ gained in a length $\ell_C$ to the energy
$\gamma m_e c^2$ lost by Compton scattering, using Eq.~\ref{lc}, yields the 
maximum electron Lorentz factor 
\begin{equation}
\gamma_C \sim \left({eEr \over m_p c^2} {r c^2 \over GM} {L_E \over 
L_{th}} \right)^{1/2}.
\label{gc}
\end{equation}
Using $E \sim v B/c \sim B (GM/r c^2)^{1/2}$ and Eq.~\ref{b} yields
\begin{eqnarray}
\gamma_C &\sim &\left({r c^2 \over GM}\right)^{1/8} \left({L_{th} \over 
L_E}\right)^{-1/2} \left({r \over h}\right)^{1/4} \left({GN_B m_e^2 \over
e^2} \right)^{1/4} \left({L_b \over L_E}\right)^{1/4} \nonumber \\ &\sim &
1.0 \times 10^4 \left({M \over M_\odot}\right)^{1/4} \left({r c^2 \over
10 GM}\right)^{1/8} \left({L_{th} \over L_E}\right)^{-1/2} \left({r \over
10 h} \right)^{1/4} \left({L_b \over L_E}\right)^{1/4},
\label{gc2}
\end{eqnarray}
where $N_B \approx 1.2 \times 10^{57}\,M/M_\odot$ is the ratio of the black
hole mass to the proton mass.  It is possible to express $\gamma_C$ in terms
of fundamental constants, dropping factors of order unity which depend on
the properties of the individual object, and noting that if $M$ equals the
Chandrasekhar mass $M_{Ch}$, then $N_B \approx (\hbar c/G m_p^2)^{3/2}$:
\begin{equation}
\gamma_C \sim \left({m_e \over m_p}\right)^{1/2} \left({e^2 \over \hbar c}
\right)^{-1/4} \left({G m_p^2 \over \hbar c}\right)^{-1/8} \left({M \over 
M_{Ch}}\right)^{1/4}.
\label{gc3}
\end{equation}

\subsubsection{Compton gamma-rays}
\label{compton}

Eq.~\ref{gc2} directly gives an upper bound on the Compton scattered photon
energy $\gamma_C m_e c^2$.  This limit is $\sim 10^{12}$ eV for typical AGN
masses and $\sim 10^{10}$ eV for typical BHXRB masses if the factors in
parentheses other than $M/M_\odot$ are of order unity.  In AGN typically
$L_{th} \ll L_E$ so that significantly more energetic electrons and Compton
scattered gamma-rays may be produced.  This explains, at least
qualitatively, the production of TeV gamma-rays in AGN such as Mrk 421
(\cite{G96}) and Mrk 501 (\cite{Q96}).  

The actual spectral cutoff depends on the spectrum of thermal photons; if 
their energy $\hbar \omega_{th} \sim m_e c^2 / \gamma_C$ the cutoff will be 
$\sim \gamma_C m_e c^2$; if $\hbar \omega_{th} < m_e c^2 / \gamma_C$ the 
cutoff will be $\sim \gamma_C^2 \hbar \omega_{th}$; if $\hbar \omega_{th} > 
m_e c^2 / \gamma_C$ the cutoff is $\sim \gamma_C m_e c^2$ but Eq.~\ref{gc2} 
then underestimates $\gamma_C$ because of the reduction in Klein-Nishina 
cross-section and the discreteness of the energy loss.  The observed spectra
of AGN and BHXRB are so complicated (it is also unclear which components are 
emitted at the small radii at which electron acceleration is assumed to 
occur) that it is difficult to be quantitative.  Visible radiation from AGN 
and soft X-rays from BHXRB place Compton scattering marginally in the 
Klein-Nishina range (note that both the black body $\hbar \omega_{th}$ and 
$m_e c^2 / \gamma_C$ scale $\propto M^{-1/4}$), but quantitative
estimates depend on the uncertain factors in parentheses in Eq~\ref{gc2}.

Because the Compton scattering power of an electron and the frequency of the
scattered photon each are proportional to $\gamma^2$, while the rate of 
energy gain in an electric field is independent of $\gamma$, there is 
expected to be a broad peak in $\nu F_\nu$ around the cutoff frequency, with
$F_\nu \propto \nu^{1/2}$ at lower frequencies (the same slope as for 
synchrotron radiation discussed in \ref{synchrotron}, for the same reasons).
In principle, the weak dependence of $\gamma_C$ on $M$, $L_{th}$ and $L_b$
in Eq.~\ref{gc2} could be tested by comparing AGN and BHXRB if $M$ could be
estimated independently of the luminosity.

The majority of the power which goes into lepton acceleration may appear
as Compton scattered gamma-rays of energy $\sim \gamma_C m_e c^2$.  This
is half $P_{rw}$ in a proton-electron wind and all of $P_{rw}$ if pairs
are accelerated.  As discussed in \ref{nonthermal}, this can far exceed
$L_{th}$ if the disc is undergoing highly supercritical accretion.  This
may explain the dominance of the emitted power by energetic gamma-rays in
some AGN.  Supercritical accretion by black holes of comparatively low mass
also permits more rapid variability than accretion at the Eddington limit by
a more massive black hole, so that, in principle, masses could be estimated
from the variability time scale.

\subsubsection{Synchrotron radiation}
\label{synchrotron}

Electrons with Lorentz factors up to that given by Eq.~\ref{gc2} radiate
synchrotron radiation in the magnetic field Eq.~\ref{b}.  Assuming an
isotropic distribution of pitch angles, the characteristic synchrotron
frequencies extend up to
\begin{equation}
\nu_{synch} \sim \left({3 \over 8 \pi^2}\right)^{1/2} \left({GM \over r 
c^2} \right)^{1/2} {L_b \over L_{th}} {r \over h} {m_e c^3 \over e^2} \sim 7 
\times 10^{22}\,{L_b \over L_{th}} \left({10 GM \over r c^2}\right)^{1/2}
{r \over 10 h}\ {\rm Hz}
\label{synch}
\end{equation}.
The usual relation between the particle distribution function and the
synchrotron spectral index for $\nu < \nu_{synch}$ would predict a spectral
index of $-1/2$, because a uniform accelerating electric field produces a
particle distribution function with energy exponent zero.  However, the low
frequency component of the synchrotron radiation function increases the 
spectral index to $-1/3$, in analogy to the predicted (\cite{K94b}) and
observed (\cite{G98,FWK00,B03}) low frequency spectra of gamma-ray bursts.

The ratio of synchrotron to Compton scattering 
powers, assuming an isotropic electron distribution, is the ratio of the 
magnetic to the thermal energy densities $U_B/U_{th}$:
\begin{equation}
{P_{synch} \over P_{Compt}} = {U_B \over U_{th}} \sim {1 \over 2} {L_b
\over L_{th}} {r \over h} \left({GM \over r c^2}\right)^{-1/2}.
\label{psynch}
\end{equation}
This is generally $\ge 1$.

Unlike the Compton cutoff, the synchrotron cutoff Eq.~\ref{synch} is
independent of the mass of the black hole, and is in that sense
``universal'', although it depends on other parameters, most notably the
ratio $L_b/L_{th}$.  Eqs.~\ref{synch} and \ref{psynch} suggest the possibility of synchrotron radiation 
with significant power up to $\sim$ GeV energies.  This is probably a great
overestimate, both of the power and of the radiation frequency, because the
assumption of an isotropic distribution of electron momenta is unlikely to
be valid.  Electrons may be effectively accelerated only parallel to the
magnetic field---a component of $E$ perpendicular to $B$ does not
effectively accelerate charged particles unless it varies at their cyclotron
frequency, unlike the nearly steady corotational electric field---and
readily lose their transverse momentum by synchrotron radiation.

Internal dissipation in the relativistic wind may be as essential to
radiation in AGN as in GRB, for in them plasma turbulence may partially
isotropize the electron distribution, making effective synchrotron radiation
possible.  If the pitch angles remain small the frequency of the synchrotron
radiation is reduced and it is emitted nearly parallel to the direction of
the electrons' motion, the magnetic field, and the Compton scattered
gamma-rays.  This can be described as relativistic bulk motion of the
electrons and their associated radiation field, and may be necessary to
avoid absorption of the photons by gamma-gamma pair production.

\subsubsection{Curvature radiation}
\label{curvature1}

The electrons also radiate curvature radiation (on the magnetic field lines
of radii of curvature $\sim r$) at frequencies up to
\begin{eqnarray}
\nu_{curv} &\sim &{1 \over 2 \pi} \left({GM \over r c^2}\right)^{5/8} \left(
{L_{th} \over L_E}\right)^{-3/2} \left({L_b \over L_E}\right)^{3/4} \left({r
\over h}\right)^{3/4} {m_e^{3/2} c^3 \over m_p^{3/4} e^{3/2} (GM)^{1/4}}
\nonumber \\
&\sim &3 \times 10^{15} \left({10 GM \over r c^2}\right)^{5/8} \left({L_{th}
\over L_E}\right)^{-3/2} \left({L_b \over L_E}\right)^{3/4} \left({r \over
10 h}\right)^{3/4} \left({M_\odot \over M}\right)^{1/4}\ {\rm Hz}. 
\end{eqnarray}

Curvature radiation is insignificant, and its power is small, when the
electron energy is limited by Compton scattering in the Thomson limit.  It
cannot cause pair production, in contrast to the case of pulsars (which
have much larger fields than do accretion discs).

\subsubsection{Pair production}
\label{pair1}

The most energetic gamma-rays (produced when the electrons Compton scatter
thermal photons) of energy $E_\gamma \sim \gamma_C m_e c^2$ may produce
electron-positron pairs by interacting with the thermal photons.  The
condition for this to occur (assuming an isotropic thermal radiation field)
\begin{equation}
E_\gamma \hbar \omega_{th} \sim \gamma_C m_e c^2 \hbar \omega_{th} > 
(m_e c^2)^2,
\label{pairth}
\end{equation}
is equivalent to the condition for the breakdown of the Thomson 
approximation to Compton scattering.  It appears to be met in AGN and BHXRB, 
taking the observed thermal spectra and assuming all factors in parentheses 
in Eq.~\ref{gc2}, except that involving the mass, are $O(1)$.  If the
thermal radiation field is assumed to be a black body then Eq.~\ref{pairth}
can be rewritten
\begin{equation}
\left({GM \over r c^2}\right)^{3/8} \left({r \over h}\right)^{1/4} \left(
{L_b \over L_{th}}\right)^{1/4} \left({e^2 \over \hbar c}\right)^{-3/4} > 1.
\end{equation}
It is clear that this condition is generally met; if (as is likely) the 
thermal spectrum is harder than that of a black body at the effective 
temperature the inequality holds even more strongly.  Pair production by 
interaction between gamma-rays produced by Compton scattering of the 
accelerated electrons and thermal photons takes the place of pair production
by curvature radiation which occurs in pulsars.  

\subsection{Electron Retardation by Compton Scattering; Klein-Nishina Case}
\label{klein}

As pointed out in \ref{compton}, Compton scattering of the accelerated
electrons by thermal photons may be at sufficiently high energy that the
Thomson cross-section is inapplicable, and the full Klein-Nishina
cross-section must be used instead.  In this case $\ell_C$ depends on the
frequency distribution of the thermal radiation as well as on its
luminosity.  Eq.~\ref{lc} is replaced by 
\begin{equation}
\ell_{KN} = {\gamma \over 3 \ln{(\gamma h \nu_{th}/m_e c^2)}} {h \nu_{th}
\over m_e c^2} {L_E \over L_{th}} {h \nu_{th} r \over GM m_p} r.
\label{lkn}
\end{equation}
The dependence on the thermal photon frequency $\nu_{th}$ is nearly
quadratic, one power coming from the reciprocal relation between the photon
number density and their frequency (at fixed $L_{th}$) and the other power
coming from the energy dependence of the Klein-Nishina cross-section.  We
have assumed a thermal radiation intensity $L_{th}/4 \pi r^2$, as would be
produced by energy dissipated at radii comparable to that of particle
acceleration.  Thermal radiation produced at greater radii (for example,
in the broad line regions of AGN) is diluted, giving a correspondingly
lower effective value of $L_{th}$.

Because of this sensitivity to $\nu_{th}$, information about the spectrum is
required to evaluate Eq.~\ref{lkn}.  For example, for near-Eddington
limited accretion onto a $10^8 M_\odot$ black hole $\ell_{KN} \gg r$ if $h
\nu_{th} = 10$ KeV, but $\ell_{KN} \ll r$ if $h \nu_{th} = 100$ eV.  A small
amount of soft radiation has a large effect.  For electrons of Lorentz
factor $\sim 10^6$, suggested in \ref{thomson}, even scattering by visible
light (below the black body spectral peak for luminous accretion onto a
supermassive black hole) or the ``blue bump'' observed in some AGN spectra
is in the Klein-Nishina regime, and the efficacy of Compton scattering in
limiting electron acceleration is sensitive to the actual spectral
distribution of the radiation.

The rate of electron energy loss is essentially proportional to the
scattering rate, because an electron loses most of its energy in a single
scattering.  The Compton scattering rate is given by $\int\,\sigma_{KN}
(F_\nu/\nu)\,d\nu$, where $\sigma_{KN} \propto \ln{(h\nu/m_e c^2)}/\nu$.  If
the thermal spectrum is self-absorbed below a frequency $\nu_{abs}$, follows
a thin bremsstrahlung or synchrotron spectrum at higher frequencies, and
$\nu_{abs} > \nu_{KN}$ where the characteristic Klein-Nishina frequency
$\nu_{KN} \equiv m_e c^2/\gamma h$, then most of the Compton energy loss is
attributable to photons of frequency $\nu \sim \nu_{abs}$ and $L_{th}$ in
Eqs.~\ref{lc}, \ref{gc} and \ref{gc2} should be replaced by the luminosity
at frequencies of this order.  If $\nu_{abs} < \nu_{KN}$ then (provided the
spectral index in the optically thin regime does not exceed 1) most of the
Compton energy loss is attributable to photons of frequency $\nu \sim
\nu_{KN}$, and $L_{th}$ should be replaced by the luminosity at frequencies
of this order.  These are reasonable rough approximations to the full
integrals of the Compton energy transfer function (\cite{BG70}) over the
radiation spectrum.

\subsubsection{Ultimate (Goldreich-Julian) limit}
\label{ultimate}

The reduction in the Compton energy losses when the bulk of the thermal
luminosity is emitted at frequencies above $\nu_{KN}$ implies that the
maximum Lorentz factor is larger than indicated in Eqs.~\ref{gc}, \ref{gc2}
and \ref{gc3}.  The available potential drop is given by \cite{GJ69} and
accelerated particles of mass $m$ reach a limiting Lorentz factor
\begin{equation}
\gamma_{GJ} \sim {e \over 2 m c^2} \left({\Omega r \over c}\right)^2 r B,
\label{ggj}
\end{equation}
in place of Eq.~\ref{gc}.  Using the na\"\i ve estimate Eq.~\ref{b} for $B$
leads to a Lorentz factor of accelerated electrons, in place of
Eq.~\ref{gc2},
\begin{eqnarray}
\gamma_{GJe} &\sim &\left({r c^2 \over GM}\right)^{-3/4} \left({r \over
h}\right)^{1/2} \left({GM m_p \over e^2}\right)^{1/2} \left({L_b \over
L_E}\right)^{1/2} \nonumber \\ &\sim &2 \times 10^{10}\ \left({M \over
M_\odot}\right)^{1/2} \left({r c^2 \over 10 GM}\right)^{-3/4} \left({r \over
10h}\right)^{1/2} \left({L_b \over L_E}\right)^{1/2}.
\label{ggj2}
\end{eqnarray}
This may be rewritten in terms of fundamental constants, in analogy to 
Eq.~\ref{gc3}:
\begin{equation}
\gamma_{GJe} \sim \left({e^2 \over \hbar c}\right)^{-1/2} \left({G m_p^2
\over \hbar c}\right)^{-1/4} \left({M \over M_{Ch}}\right)^{1/2}.
\label{ggj3}
\end{equation}
For a black hole of mass $M \sim 10^8 M_\odot$, as expected for AGN,
electrons may be accelerated to energies $\sim 10^{20}$ eV.  For black holes
of stellar mass the limiting electron energy is $\sim 10^{16}$ eV.  If this
energy is achieved in microquasars, they may produce (by Compton scattering)
gamma-rays approaching this energy.

\subsubsection{Curvature radiation}
\label{curvature2}

The extreme energies of Eqs.~\ref{ggj}, \ref{ggj2} and \ref{ggj3} are
possible only in the limit in which Compton drag is negligible, either
because $L_{th}$ is small or because $\nu_{th}$ is large (Eq.~ref{lkn}).  If
such extreme electron energies are achieved then even the curvature
radiation may be powerful and energetic.  Equating the accelerating power
$eEc \sim evB$ to the curvature radiation loss $2 e^2 c \gamma^4 / (3 r^2)$
and using Eq.~\ref{b} for $B$ yields a limiting Lorentz factor
\begin{eqnarray}
\gamma_{curv} & \sim & \left({r c^2 \over GM}\right)^{1/16} \left({r \over
h} \right)^{1/8} \left({L_b \over L_E}\right)^{1/8} \left({r \over r_e}
\right)^{3/8} \left({m_p \over m_e}\right)^{1/8} \nonumber \\ & \sim & 4
\times 10^7 \left({r c^2 \over 10 GM}\right)^{7/16} \left({r \over 10 h}
\right)^{1/8} \left({L_b \over L_E}\right)^{1/8} \left({M \over M_\odot}
\right)^{3/8},
\end{eqnarray}
where $r_e \equiv e^2 / m_e c^2 = 2.82 \times 10^{-13}$ cm is the classical
electron radius.  Just as for pulsars, if Compton drag is negligible the
electron energy will be limited by curvature radiation losses, and most of
the power of electron acceleration will appear as curvature radiation.

The characteristic frequency of curvature radiation is then
\begin{eqnarray}
\nu_{curv} & \sim & {1 \over 2 \pi} \left({10 GM \over r c^2}\right)^{1/8}
\left({r \over 10 h}\right)^{1/4} \left({L_b \over L_E}\right)^{1/4}
{c^3 \over 10 GM_\odot} \left({10 GM_\odot \over r_e c^2}\right)^{3/4}
\left({m_p \over m_e}\right)^{1/4} \nonumber \\ & \sim & 6 \times 10^{18}
\left({10 GM \over r c^2}\right)^{1/8} \left({M_\odot \over M}\right)^{1/4}
\left({r \over 10 h}\right)^{1/4} \left({L_b \over L_E}\right)^{1/4}\
{\rm Hz}.
\label{curv}
\end{eqnarray}
This frequency, typically hard X-rays for stellar-mass objects and very soft
X-rays for AGN, is not a prominent feature of their spectra.  This implies
that curvature radiation is not usually the dominant electron energy loss
mechanism in these objects, and supports the customary assumption that
Compton loss is dominant.

\subsubsection{Pair production}
\label{pair2}

Very energetic electrons lose more energy on a background radiation field
by pair production ($\gamma e^- \to e^-e^-e^+$) than by Compton scattering
(\cite{MMB86}).  These authors find that this is the case only far into
the Klein-Nishina regime, when $\gamma_e > 550 m_ec^2/h\nu_{th}$.  Using
their results, we obtain a limiting Lorentz factor
\begin{equation}
\gamma_{e,pair} \sim 6 \times 10^{11} \left({rc^2 \over GM}\right)^{0.98}
\left({r \over h}\right)^{0.65} \left({L_b \over L_E}\right)^{0.65}
\left({L_E \over L_{th}}\right)^{1.30} \left({h\nu_{th} \over m_e c^2}
\right)^{1.60} \left({M \over M_\odot}\right)^{0.65}.
\end{equation}
Of course, if this $\gamma_{e,pair}$ exceeds the Goldreich-Julian limit
Eq.~\ref{ggj} then radiation drag is negligible and the actual limit is
$\gamma_e < \gamma_{GJe}$.

Because of the sensitivity of $\gamma_{e,pair}$ to $\nu_{th}$ even a small
amoung of soft radiation has a large effect on the limiting electron and
positron energy, and hence on the photon energy of Compton gamma rays and of
curvature radiation produced by these light leptons.  This is the same
difficulty encountered when Compton drag in the Klein-Nishina limit is
dominant.   Without a detailed geometrical and spectral model it is not
possible to estimate $\gamma_{e,pair}$ even to order of magnitude.

\subsection{Proton acceleration}
\label{protonacc}

If protons are accelerated their energy may be radiated as high energy
gamma-rays following collisional pion production, either directly from
$\pi^0$ decay or by Compton scattering of $e^\pm$ produced by decay of
$\pi^\pm$.  In the latter case neutrinos are also produced, with power and
energy comparable to that of the gamma-rays.  Alternatively, the proton 
kinetic energy may be degraded and coupled to electrons in a collisionless
shock, as is generally assumed to occur in gamma-ray bursts, and then
radiated by Compton scattering or by synchrotron radiation.  The proton
kinetic energy may much exceed the values given by Eqs.~\ref{gc}, \ref{gc2}
and \ref{gc3}, and may approach the limiting energies given in
\ref{ultimate}.  Gamma-rays (and neutrinos) following from pion production
will have energies approaching a tenth of the proton energies.  On the other
hand, proton synchrotron radiation will occur at much lower energies than
electron synchrotron radiation, and is generally negligible.

If there were no pair production, the accelerated plasma would consist of 
protons (and nuclei) and electrons.  Even in the presence of pair 
production, protons may be accelerated along with the positrons.  This is 
important in intense sources of thermal radiation, such as BHXRB and AGN, 
because proton-photon scattering is negligible below the pion production 
threshold.  Even above threshold, the effective energy loss cross-section 
(\cite{G66}) is a fraction $f \sim 10^{-4}$ of the Thomson cross-section.  
This permits accretion discs around black holes in AGN and BHXRB to be 
efficient proton accelerators (as has previously been discussed by 
\cite{LB69,KE86,K91} in other models).  Very high energy gamma-rays may then
result from photoproduction of $\pi^0$ by thermal radiation.  Collisions of
protons with nucleons make pions, leading to high energy radiation directly
from $\pi^0$ decay and indirectly by Compton scattering of the very
energetic $e^\pm$ produced by $\pi^\pm \to \mu^\pm \to e^\pm$
(\cite{K91,DL97}).  Similarly, very high energy neutrinos are produced in
the pion and muon decays.

The energy loss length of a proton is, in analogy to Eq~\ref{lc},
\begin{equation}
\ell_p \approx {1 \over \gamma f} {L_E \over L_{th}} {r c^2 \over GM} r.
\end{equation}
Unlike Thomson scattering, this process has an energy threshold determined
by the $\pi^0$ rest mass of 135 MeV.  If the thermal spectrum consists of 
visible light, the acceleration of protons is not restrained until $\gamma_p
\sim 10^8$.  The result analogous to Eq.~\ref{gc2} for the limiting Lorentz
factor of protons slowed by photopion production in a thermal radiation
field is
\begin{eqnarray}
\gamma_p &\sim &\left({r c^2 \over GM}\right)^{1/8} \left({L_{th} \over L_E}
\right)^{-1/2} \left({r \over h}\right)^{1/4} \left({GNm_e^2 \over e^2}
\right)^{1/4} \left({L_b \over L_E}\right)^{1/4} f^{-1/2} \nonumber \\ &\sim
&1.0 \times 10^6\ \left({M \over M_\odot}\right)^{1/4} \left({r c^2 \over 10
GM} \right)^{1/8} \left({L_{th} \over L_E}\right)^{-1/2} \left({r \over 10
h} \right)^{1/4} \left({L_b \over L_E}\right)^{1/4}.
\end{eqnarray}
For an AGN with $M \sim 10^8 M_\odot$ this yields a limiting $\gamma_p \sim 
10^8$, approximately the threshold at which $\pi^0$ photoproduction begins.
Just as for pair production, the product of $\gamma_p$ and the black-body 
thermal photon energy is independent of $M$, so that Eq.~\ref{curv} may be
(barely) applicable at all $M$.

\subsubsection{Proton Radiation}
\label{protonrad}

These possible $\sim 10^{17}$ eV protons in AGN could produce $\sim
10^{16}$ eV photons from photoproduced $\pi^0$, but these are not observable
at the greatest distances because of pair production on the microwave
background radiation.  However, at closer distances these gamma-rays may be
observable.  In BHXRB, perhaps including microquasars, the corresponding
proton energies are $\sim 10^{15}$ eV, which may produce photons of $\sim
10^{14}$ eV.

\subsubsection{Underluminous black holes as sources of UHE cosmic rays?}
\label{underluminous}

If most of the thermal radiation is ineffective at slowing the protons
because it is below the photopion threshold (in the protons' frame) or if
$L_{th}$ is small, then, just as for electrons in the Klein-Nishina case, it
may be possible for protons to approach the limiting Lorentz factor
Eq.~\ref{ggj}.  The low luminosity black hole powering Sgr A* in our
Galactic Center demonstrates that black holes with $L_{th} / L_E \ll 1$ may
exist even in the presence of sufficient mass to support high accretion
rates, and they would generally be undetectable at extragalactic distances.
Ratios $L_{th}/L_E$ even lower than that of Sgr A* may occur.  For
parameters appropriate to supermassive black holes the proton energies of
$\sim 10^{20}$ eV, and energies of heavier nuclei greater by a factor of
$Z$, may be sufficient to explain the highest energy cosmic rays.

\subsubsection{Stellar-mass black holes as TeV sources?}
\label{stellarmass}

Just as underluminous supermassive black holes may accelerate protons and
nuclei to ultra-high energies, underluminous stellar-mass black holes may
accelerate them to energies sufficient to be sources of TeV radiation (if
the hadrons collide with other hadrons to make $\pi^0$ or $\pi^\pm$ whose
decay electrons undergo Compton scattering).  Such objects, resembling
blazars (rather than isolated neutron stars, the more conventional
explanation) might be the origin of the unidentified Galactic TeV sources
(\cite{U06}).

\section{Are there shocks in GRB?}
\label{shocks}

In most models of relativistic astronomical flows it is assumed that the
fluids behave as if they were collisional, even though the single-particle
collision lengths are generally orders of magnitude greater than the spatial
extent of the fluids involved.  Collective collisionless processes
(beam-driven electromagnetic or longitudinal two-stream instabilities) are
assumed to lead to momentum transfer on much shorter scales, so that the
flows can be described by the hydrodynamic equations.  Kinematic constraints
require some such interaction in order to convert the kinetic energy of
relativistic motion to radiation.  In particular, in a hydrodynamic
supersonic flow the interaction takes the form of shocks.  Such shocks are
assumed in nearly all models of GRB.

GRB are the only identified objects near whose central engines ${\cal I}
\gg 1$.  The observed radiation is produced at distances several orders of
magnitude greater (typical estimates are ${\cal O}10^{15}$ cm), at which
${\cal I} \ll 1$.  Under these conditions nonthermal particle acceleration
is possible, and in fact is required by their observed nonthermal spectra.
Further development of models requires investigation of the conditions under
which thermal flows lead to the acceleration of nonthermal particles.  Shock
processes have been extensively investigated, but we should also ask if
shocks (hydrodynamic discontinuities between two fluids each with
thermodynamic equilibrium distribution particle functions, though generally
transparent and with photon energy densities many orders of magnitude less
than Planckian at the matter temperature) form at all.  
The kinematic constraints cannot be avoided, but it may be that the
hydrodynamic assumption is invalid.  It has been demonstrated
(\cite{FMN96,SP97}) that the complex multipeaked temporal structure of GRB
can only be explained, given the hydrodynamic assumption, by internal 
shocks.  External shocks may be associated with, or even defined as, the
origin of afterglows (\cite{K98}).  Yet elementary kinematics
(\cite{K97,KP97,FP06}) shows that, unless the ratios of Lorentz factors are
very large, the energetic efficiency of internal shocks is low, generally
no more than 10--20\%.  More efficient radiation would require the emission
of a shell of low Lorentz factor followed by one of much higher Lorentz
factor.  It is implausible that this occurs regularly, but any deviation
from this sequence (for example, a random distribution of Lorentz factors
and of proper masses) would imply low efficiency.  This would increase the
energetic requirements of GRB and (perhaps worse) would raise the question
of why the remaining energy does not appear as afterglows many times more
energetic than GRB themselves.

One possible escape from this ``Efficiency Crisis'' (\cite{I05}) is to
reject the assumption of hydrodynamic flow.  Recall the reason why an
external shock (on external matter initially at rest or moving 
non-relativistically) cannot explain complex substructure: After the
external matter is hit by a shell of relativistic debris, it must be
accelerated in a shock to relativistic speeds (Lorentz factor $> 100$) to
explain the observed properties of GRB emission (spectrum and avoidance of
gamma-gamma pair production).  Then the external matter is no longer
available as a static or non-relativistically moving target for the impact
of a second shell, and the duration of emission of the first subpulse (from
impact of the first shell) will overlap that of a second subpulse, even if
a second shell is emitted cleanly separated from the first shell
(\cite{FMN96,SP97}).

This conclusion may be avoided if the first relativistic shell interacts
only with a fraction of the matter at each point surrounding it, thus
satisfying the kinematic constraints while violating the hydrodynamic
assumption that the target behaves as a single fluid with a single
(relativistic) equilibrium particle distribution function.  It is possible
that the collective plasma processes that mediate the interaction couple, at
any time, only to a small fraction of the matter (for example, to a narrow
subrange of the velocity distribution) in the external medium.  This
is, at least qualitatively, consistent with the intermittent and fluctuating
nature of most observed plasma-physical instabilities, both in the
laboratory and in Nature.

This hypothesis suggests that as the density of target matter is gradually
eroded by the passage through it of successive shells of relativistic debris
the parameters of the accelerated particles and the properties of their
radiation may also change.  Although the details of these changes are
unknown (because the collective interactions between debris and target are
not understood from first principles), this does suggest a progressive
evolution of the characteristics of GRB emission through the pulse,
consistent with the usual (but not universal) observations that the emission
gradually softens.

Unfortunately, this is not an unambiguous test of the hypothesis because the
parameters and properties of the central engine are also changing as its
mass is gradually accreted on to a central black hole, and this can be an
alternative explanation of any progressive changes through GRB pulses.  The
statistics of sub-pulses may be different in the internal shock and
non-shock models, but in both cases depend on the unknown statistics of
the activity of the central engine.

\section{Discussion}
\label{discussion}

It is apparent from comparing the results of \ref{thomson} to those of
\ref{klein} that the actual particle energies achieved depend sensitively on
the flux and spectral distribution of any soft radiation in the acceleration
region.  The power and spectral distribution of radiation produced by
interaction (Compton scattering, pion photoproduction) with this soft
radiation also depend on the properties of the soft radiation, partly
directly, and partly because it affects the distribution of energy of the
energetic particles.  A small amount of dense nonrelativistic plasma,
producing a black body or thin bremsstrahlung flux may have a large effect
on the more energetic radiation of an accreting black hole, even though the
thermal radiation is comparatively insignificant in power.  The kinematic
reason for this is similar to that encountered in the study of GRB external
shocks, in which a proper mass $E/\Gamma^2 c^2$ at rest is sufficient to
dissipate inelastically a kinetic energy $E$ in matter moving with Lorentz
factor $\Gamma$.

In the present problem $\Gamma$ may be as large as $\sim 10^{11}$ for at 
least a few of the energetic particles, and generally $\Gamma \gg 1$ by 
orders of magnitude, so that a tiny amount of isotropic thermal radiation
(substituting for mass at rest) may have a large effect on the energetic
particles if the interaction cross-sections are sufficient.  This
extraordinary sensitivity implies that predictive quantitative models are
difficult to build.  For example,  for supermassive black holes possible 
objects range from classical quasars (with $L_{th} \sim L_E \sim 10^{46}$
erg/s) to sources of the most energetic cosmic rays with almost no 
electromagnetic luminosity at all.

Emergent spectra depend on such poorly understood and
essentially unpredictable variables as the flux and angular distribution of
thermal radiation (and hence on the location and parameters of thermal gas)
and on temporally fluctuating collective interactions between essentially 
collisionless interpenetrating streams of matter with relativistic
relative velocities.  Qualitatively, this sensitivity is consistent with
the complex time structure observed in AGN, BHXRB and GRB, but makes it
difficult to predict the emergent spectra.  However, linear polarization
is a general property of most nonthermal radiation processes (most
familiarly, synchrotron radiation), and is predicted if the X-ray emission
of BHXRB is nonthermal in some of their states.

AGN may show QPO, just as suggested for GRB (\cite{K97}), but with typical
periods of order hours to days, depending on the black hole masses and
angular momenta (\cite{ST83}).  No such QPO have been found in the extensive
body of visible light data on AGN.  This may be explained if the visible
light is produced far from the central object or by thermal radiation from a
nearly axisymmetric gas disc.  It may be more fruitful (though more
difficult observationally) to search for QPO in the energetic gamma-rays
produced as particles are accelerated along magnetic field lines closer to
the rotating disc.  It is the magnetic field that would be expected to show
the greatest deviation from axisymmetry, as in pulsars.

The considerations of this paper have led to the prediction of two novel
kinds of objects: 1. Young, high field, rapidly rotating pulsars (with
$\mu^2 \omega^6 \gg 12 \pi c^5 I_{char} \approx 10^{73}$ erg cm$^3$/s$^6$),
that produce a black-body equilibrium pair gas wind rather than
the few but very energetic particles produced by radio pulsars.  Pulsars
satisfying this condition on $\mu^2 \omega^6$ may resemble a SGR in permanent
outburst for a lifetime $\sim {\cal I} \omega^4/(8 \pi c^2 I_{char}) \sim
3 \times 10^{-7} \omega^4$ s$^5$, where $\cal I$ is the moment of inertia.
The minimum luminosity of such an object is $4 \pi r^2 I_{char} \sim
10^{42}$ erg/s, so they would be detectable (following supernovae or quiet
stellar collapse) as rapid ($\sim 1$ KHz) periodic X-ray sources at
distances approaching that of the Virgo cluster ($\approx 20$ Mpc).
2. Supermassive black holes with most of their luminosity in ultra high
energy particles, and perhaps not recognizable as AGN.  If these are the
source of UHE cosmic rays, their arrival directions will be clumped (subject
to propagation effects) and perhaps correlated with point sources of UHE
neutrinos and gamma-rays.

\section*{Acknowledgements}

I thank T. Piran, M. A. Ruderman and R. Sari for discussions, Washington
University for the grant of sabbatical leave, and the Hebrew University for
hospitality and a Forchheimer Fellowship.
\appendix
\section{Conditions for Particle Acceleration}
The central mystery of AGN is why so much of their
accretional power appears as particle acceleration.
In order to accelerate energetic particles, necessary both for incoherent
emission (as in AGN and radio sources) and for coherent emission (as in 
pulsars and FRB, if these particles drive a plasma instability), it is
necessary that they gain energy from an electric field faster than they lose
it to interaction with ambient plasma (by ``Coulomb drag'') \citep{A09}.
For a relativistic electron the ratio of these quantities defines an
acceleration parameter
\begin{equation}
	\label{A}
	A \approx {E m_e c^2 \over 4 \pi e^3 n \ln{\Lambda}},
\end{equation}
where $E$ is the electric field, $n$ the plasma particle density and
$\Lambda$ is $2 m_e c^2/I$ with $I$ the ionization potential in a neutral
medium or $m_ec^2/\hbar \omega_p$ in a plasma.  $\ln{\Lambda} \approx 20$
in most astronomical environments, and is insensitive to their parameters.
$A > 1$ is a necessary condition for particle acceleration.

For the midplane of an accretion disc, adopting the simple scaling model
presented here and taking
\begin{equation}
	E \sim {v_{orb} \over c}B \sim \sqrt{4\pi L \over L_E}
	\left({GM \over rc^2}\right)^{5/4} \left({r \over h}\right)^{1/2}
	\left({c^4 \over GM\kappa}\right)^{1/2},
\end{equation}
where $L$ is the accretional luminosity, $L_E$ the Eddington luminosity,
$r$ the distance from the central mass $M$, $h$ the local disc thickness
and $\kappa$ the opacity (from the definition of $L_E$), and using the same
scaling model for the density
\begin{equation}
	n \sim {L \over L_E}\sqrt{rc^2 \over GM}{1 \over \alpha} {3 \over
	4 \pi h r_e^2},
\end{equation}
where $\alpha$ is the dimensionless viscosity parameter of the disc and
$r_e$ the classical electron radius.

Combining these expressions
\begin{equation}
	\label{AAGN}
	A \sim {\sqrt{4\pi}m_e c^2 \over e^3\ln{\Lambda}} \left({L \over
	L_E}\right)^{-1/2} \left({GM \over rc^2}\right)^{7/4} \left({r
	\over h}\right)^{-1/2}\left({c^4 \over GM\kappa}\right)^{1/2}
	{r r_e^2 \alpha \over 3}.
\end{equation}
This is generally $\gg 1$, but is not realistic because in the generally
applicable magnetohydrodynamic regime $E \ll (v/c) B$.

The low density corona of an accretion disc, where magnetohydrodynamics may
not be applicable, offers a more favorable environment for the acceleration
of energetic particles.  The scaling $A \propto M^{1/2}$ in Eq.~\ref{AAGN},
with $r \propto M$ and the other factors either dimensionless and of order
unity or physical constants, may explain why AGN are proportionally more
efficient accelerators of energetic particles (their relativistic jets and
double radio lobes representing much of their total accretional power) than
stellar mass black holes (most of whose accretional power is emitted as
thermal X-rays).

\end{document}